\documentclass[10pt,conference]{IEEEtran}

\def \arXiv {}

\ifx \arXiv \undefined
\else
\pdfoutput=1
\fi

\usepackage[utf8]{inputenc}
\usepackage{acronym}
\usepackage{amsmath,amssymb,amsfonts}
\usepackage{bigints}
\usepackage[bookmarks=false]{hyperref}
\usepackage{cleveref}
\usepackage{bm}
\usepackage{nicefrac}
\usepackage{siunitx}
\usepackage{graphicx}
\usepackage{textcomp}
\usepackage{xcolor}
\usepackage{caption}
\usepackage{subcaption}
\usepackage{cite}
\usepackage{url}
\usepackage{nicefrac}

\acrodef{BNC}{bacterial nanocellulose}
\acrodef{CBD}{cannabidiol}
\acrodef{CDD}{controlled drug delivery}
\acrodef{IoBNT}{Internet of Bio-Nano-Things}
\acrodef{MC}{molecular communications}
\acrodef{MSE}{mean squared error}
\acrodef{pdf}{probability density function}
\acrodef{PLGA}{poly(lactic-co-glycolic acid)}
\acrodef{PMP}{PLGA-based microparticles}
\acrodef{RV}{random variable}
\acrodef{Rx}{receiver}
\acrodef{Tx}{transmitter}

\newcommand{\Di}{\ensuremath{D_{\mathrm{i}}}}
\newcommand{\Dihat}{\ensuremath{\hat{D}_{\mathrm{i}}}}
\newcommand{\Ditilde}{\ensuremath{\tilde{D}_{\mathrm{i}}}}
\newcommand{\Do}{\ensuremath{D_{\mathrm{o}}}}
\newcommand{\dt}{\ensuremath{\,\mathrm{d}t}}
\newcommand{\dtau}{\ensuremath{\,\mathrm{d}\tau}}
\newcommand{\dtheta}{\ensuremath{\,\mathrm{d}\theta}}
\newcommand{\dx}{\ensuremath{\,\mathrm{d}x}}
\newcommand{\Exp}[1]{\mathbb{E}\left\lbrace #1 \right\rbrace}
\newcommand{\dint}[2]{\int\limits_{#1}^{#2}}
\newcommand{\omtil}{\ensuremath{\tilde{\omega}}}
\newcommand{\Ri}{\ensuremath{R_{\mathrm{i}}}}

\renewcommand{\baselinestretch}{0.96}

\newtheorem{theorem}{Theorem}
\newtheorem{corollary}{Corollary}
    
\makeatletter
\long\def\@makecaption#1#2{\ifx\@captype\@IEEEtablestring%
    \footnotesize\begin{center}{\normalfont\footnotesize #1}\\
        {\normalfont\footnotesize\scshape #2}\end{center}%
    \@IEEEtablecaptionsepspace
    \else
    \@IEEEfigurecaptionsepspace
    \setbox\@tempboxa\hbox{\normalfont\footnotesize {#1.}~~ #2}%
    \ifdim \wd\@tempboxa >\hsize%
    \setbox\@tempboxa\hbox{\normalfont\footnotesize {#1.}~~ }%
    \parbox[t]{\hsize}{\normalfont\footnotesize \noindent\unhbox\@tempboxa#2}%
    \else
    \hbox to\hsize{\normalfont\footnotesize\hfil\box\@tempboxa\hfil}\fi\fi}
\makeatother

\begin{document}
\title{Microparticle-based Controlled Drug\\ Delivery Systems: From Experiments to\\ Statistical Analysis and Design}

\author{
    \IEEEauthorblockN{Sebastian Lotter\textsuperscript{*},
     Tom Bellmann\textsuperscript{\textdagger},
     Sophie Marx\textsuperscript{\textdagger},
     Mara Wesinger\textsuperscript{\textdagger},
     Lukas Brand\textsuperscript{*},\\
     Maximilian Sch\"afer\textsuperscript{*},
     Dagmar Fischer\textsuperscript{\textdagger}, and
     Robert Schober\textsuperscript{*}}\\
    \vspace*{-4mm}
    \IEEEauthorblockA{
        \textit{\textsuperscript{*}Institute for Digital Communications, Friedrich-Alexander-Universität Erlangen-N\"urnberg (FAU)}, Erlangen, Germany\\
        \textit{\textsuperscript{\textdagger}Division of Pharmaceutical Technology and Biopharmacy, FAU,} Erlangen, Germany
    }
    \vspace*{-8mm}
}

\maketitle

\begin{abstract}
\Ac{CDD}, the controlled release and delivery of therapeutic drugs inside the human body, is a promising approach to increase the efficacy of drug administration and reduce harmful side effects to the body.
\Ac{CDD} has been a major research focus in the field of \ac{MC} with the goal to aid the design and optimization of \ac{CDD} systems with communication theoretical analysis.
However, the existing studies of \ac{CDD} under the \ac{MC} framework are purely theoretical, and the potential of \ac{MC} for the development of practical \ac{CDD} applications remains yet to be shown.
This paper presents a step towards filling this research gap.
Specifically, we present a novel \ac{MC}-based model for a specific \ac{CDD} system in which drugs are embedded into microparticles and released gradually towards the target site.
It is demonstrated that the proposed model is able to faithfully reproduce experimental data.
Furthermore, statistical analysis is conducted to explore the impact of the microparticle size on the drug release.
The presented results reveal the sensitivity of the drug release to changes in the microparticle size.
In this way, the proposed model can be used for the design of future microparticle-based \ac{CDD} systems.
\end{abstract}

\acresetall

\section{Introduction}
\label{sec:introduction}
{\em \Ac{MC}} denotes a recently emerging communication paradigm that considers the transmission of information by means of chemical information carriers.
Specifically and in contrast to conventional communication based on electromagnetic waves, {\em molecules} are used to transmit information in \ac{MC}.
The inspiration for \ac{MC} hereby stems from natural \ac{MC} systems, in which biological entities like cells and organs encode and exchange information using molecules.
\Ac{MC} is the endeavour to leverage this natural form of communication for in-body applications that require synthetic communication \cite{akyildiz15}.

One important example application in this context is {\em \ac{CDD}}.
In \ac{CDD}, therapeutic drugs shall be deployed to a target site inside the human body such that side effects on the rest of the body are minimized.
Under the conceptual framework of \ac{MC}, the \ac{CDD} scenario resembles a communication system where an engineered \ac{Tx}, the drug releasing device, emits a signal, the drug molecules, that propagates over a physical channel, e.g., the cardiovascular system, towards the \ac{Rx}, i.e., the target site.
The \acp{Rx} in such systems are typically predetermined by the specific application.
Accordingly, the communication engineering challenge in developing \ac{CDD} systems resides mainly in the \ac{Tx} design.

In this paper, we focus on one specific \ac{CDD} system, in which drug molecules are delivered in a controlled manner from a wound dressing on the human skin to the wound.
The challenge in designing such systems is to guarantee a particular drug supply rate to the wound during the time the wound dressing is applied.
The specific system under consideration is a wound dressing containing drug-loaded polymer microparticles.
In this system, drug release occurs in a two-step process.
In a first step, drug molecules are released from the microparticles into the aqueous matrix of the wound dressing itself.
Then, in the second step, the drug molecules migrate from the wound dressing into the wound.
There are several tunable parameters in this system, such as the size of the microparticles, which can be exploited to realize a specific drug supply rate at the target site, i.e., the wound.
In order to understand the possibilities and limits of the considered system in terms of the achievable drug release profiles, mathematical modeling, specifically under the \ac{MC} framework, can be a natural and valuable tool.

\Ac{CDD} has been a focus of \ac{MC} research for more than ten years and early progress in applying \ac{MC} in the context of \ac{CDD} has been summarized in several survey papers \cite{felicetti16a,chudeokonkwo17}.
Most of the recent works on \ac{CDD} in \ac{MC} focus on the development of communication theoretical models to aid the design of efficient drug delivery systems \cite{zhao2021ReleaseRateOptimization,shrivastava2020DmoleNovelTransreceiver}.
Besides, there are some works that consider the physics-based modeling of specific \ac{MC} system components such as the \ac{Tx} \cite{rudsari2021TargetedDrugDelivery,schaefer2022ChannelResponsesMolecule}.
Most \ac{MC}-based \ac{CDD} models are {\em bottom-up} models, i.e., starting from a set of assumptions about its components, the behavior of the entire system is derived in an {\em inductive} manner.
The power of this approach lies in the fact that the impact of physically meaningful system parameters, such as particle sizes or reaction rate constants, is reflected in the model.
The drawback, however, is that bottom-up models either rely on highly idealized assumptions regarding the considered system, e.g., with respect to the relevant chemical processes, or entail high-dimensional parameter spaces.
In both cases it is difficult or even impossible to validate the models using experimental data and, consequently, their relevance for practical system design remains yet to be shown.

On the other hand, drug release models derived by experimentalists follow a different philosophy.
Here, mostly {\em top-down} models are considered and {\em deductive} reasoning is applied.
For example, a popular top-down model to describe experimentally observed drug release from microparticles is the so-called {\em Ritger-Peppas} model.
In this model, the relative drug release is modeled as
\begin{equation}
    \frac{M_t}{M_{\infty}} = kt^n,\label{eq:ritger-peppas}
\end{equation}
where $M_t/M_{\infty}$ denotes the relative amount of released drug at time $t$, and $k>0$ and $n>0$ are fitting parameters.
Based on the value of $n$ obtained from parameter fitting to experimental data, the dominant release kinetics can be {\em deduced}, e.g., $n \approx 1/2$ would correspond to diffusion-dominated drug release \cite{ritger1987SimpleEquationDescription}.
The drawback of \eqref{eq:ritger-peppas} and other top-down models (see \cite{pourtalebijahromi2020ComparisonModelsAnalysis} for a review) is that they are not well-suited for system design, since they do not reflect changes in the values of the physical system parameters.

In this paper, we propose, validate, and study a novel \ac{MC}-based model for the \ac{CDD} from microparticles embedded into a specific type of wound dressing.
In contrast to existing models, the proposed model integrates both {\em descriptive and normative} features.
In particular, we demonstrate that the model is capable of faithfully reproducing experimental data, achieving higher accuracy than benchmark top-down models with the same number of fitting parameters.
Furthermore, we present a statistical analysis that shows how the proposed model can be exploited to deduce insights for system design.
The presented results specifically provide novel insights on the role that the microparticle size can play in tailoring and optimizing the considered \ac{CDD} system.

The remainder of the paper is organized as follows.
In Section~\ref{sec:system_model}, the considered experimental \ac{CDD} setup and the analytical system model are introduced.
Section~\ref{sec:analysis} presents the statistical analysis of the proposed model.
In Section~\ref{sec:results}, numerical results are presented, and Section~\ref{sec:conclusion} concludes the paper with a brief summary and an outlook towards future research directions.

\section{System Model}
\label{sec:system_model}
\subsection{Experimental Setup}\label{sec:system_model:exp}

In the \ac{CDD} system under consideration, drug molecules shall be released in a controlled manner from a wound dressing to the wound.
When realized and applied in practice, such \ac{CDD} systems can, for example, enable the anti-microbial treatment of a burn wound during the healing process \cite{poetzinger2017BacterialNanocelluloseFuture}.
Specifically, we consider the release of {\em \ac{CBD}}, a highly lipophilic drug, from a {\em \ac{BNC} fleece}.
\Ac{BNC} is a biological material produced by bacteria \cite{iguchi2000BacterialCelluloseMasterpiece} and \ac{BNC} fleeces are ideally suited as wound dressings due to their high water content, biocompatibility, biodegradability, and thermal and mechanical stability.
To integrate the lipophilic \ac{CBD} molecules into the aqueous environment inside the \ac{BNC} fleece, they are encapsulated into degradable microparticles synthesized from {\em \ac{PLGA}} polymers, i.e., carboxylic acid-based biopolymers \cite[Ch.~7]{kalia2011BiopolymersBiomedicalEnvironmental}.
The \ac{PMP} are synthesized {\em in situ} \cite{bellmann2023SituFormationPolymer}, i.e., inside the \ac{BNC} fleece, and hence may be subject to different synthesis conditions depending on the microstructure of the respective \ac{BNC} fleece.

\begin{figure}
    \centering
    \includegraphics[width=.24\textwidth]{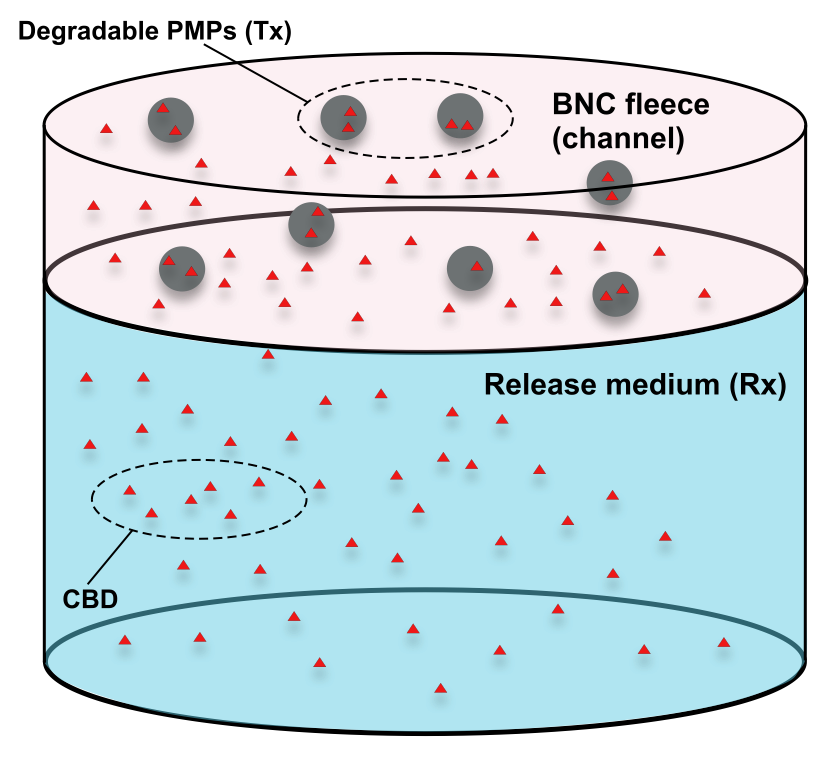}
    \caption{Schematic illustration of the considered experimental setup. Drug molecules (red triangles) are released from degradable microparticles (grey disks) into the wound dressing (top part of the cylinder) before they migrate into the release medium (bottom part of the cylinder).}
    \label{fig:system_model}
    \vspace*{-6mm}
\end{figure}

Fig.~\ref{fig:system_model} illustrates the considered \ac{CDD} system schematically.
In the experiment, the \ac{BNC} fleece containing the \ac{CBD}-loaded \acp{PMP} is placed on top of a release medium.
After the \acp{PMP} start to degrade, \ac{CBD} molecules are released into the \ac{BNC} fleece from which they further migrate into the release medium.
The release medium is sampled by the experimenter at predefined time instants to determine the \ac{CBD} release profile over time.

A complete specification of the considered experimental setup is presented in \cite{bellmann2023SituFormationPolymer}.
From this specification, the following conclusions relevant for the further modeling can be drawn.
\begin{enumerate}
    \item Both the release of \ac{CBD} from the \acp{PMP} into the \ac{BNC} and from the \ac{BNC} into the release medium are driven by the corresponding concentration gradients.
    \item The \ac{BNC} fleece and the release medium can be considered perfect sinks, i.e., the migration of \ac{CBD} molecules in reverse direction is negligible for both release processes.
    \item There is no diffusion barrier to be considered between the \ac{BNC} fleece and the release medium.
    \item The \acp{PMP} are uniformly distributed inside the \ac{BNC} matrix and the release medium is well-stirred.
\end{enumerate}

The \acp{PMP} are biodegradable and start to erode upon contact with the aqueous medium inside the \ac{BNC} fleece.
Interestingly, manipulating this process by adjusting the properties of the \acp{PMP}, e.g., their sizes, presents a degree of freedom by which the drug release can be controlled.
The results presented later in this paper provide quantitative insights into how this can be achieved.

\subsection{MC System Model}

In this section, we match the components of the \ac{CDD} system described in the previous section to communication system components and provide mathematical models for each of the components.
In this way, an \ac{MC}-based end-to-end model of the \ac{CBD} release is obtained.

\subsubsection{Transmitter}

Within the \ac{MC} framework, the \acp{PMP} and the (normalized) amount of \ac{CBD} released from the \acp{PMP} into the \ac{BNC} fleece per unit time, denoted in the following as $x(t)$, correspond to the \ac{Tx} and the transmit signal, respectively.
Several studies \cite{siepmann2005HowAutocatalysisAccelerates,klose2008PlgaBasedDrug,casalini2014MathematicalModelingPlga} have accumulated experimental evidence that drug release from \acp{PMP} can be modelled as a diffusion process.
Hence, $x(t)$ is obtained as the solution to the spherical diffusion equation with Dirichlet boundary condition and uniform initial concentration as follows\footnote{In the proposed model, the total drug concentration is normalized to $1$, since, similar to \eqref{eq:ritger-peppas}, we study the {\em relative} drug release.} \cite{crank79,siepmann2005HowAutocatalysisAccelerates}
\begin{equation}
    x(t) = \frac{6 \Di}{\Ri^2} \sum_{n=1}^{\infty} \exp\left(-\frac{\Di n^2 \pi^2}{\Ri^2} t \right), \qquad t \geq 0,
    \label{eq:def:n_t}
\end{equation}
where $\Ri$ and $\Di$ denote the radius of the \acp{PMP} in \si{\milli\meter} and the {\em effective} diffusion coefficient in \si{\milli\meter\squared\per\hour} of the \ac{CBD} molecules inside the \acp{PMP}, respectively.

It is important to note that the ``diffusion'' in this context does not exclusively arise from Brownian motion inside the \acp{PMP}.
Instead, it results from several bio-physical processes including Brownian motion, but also other effects such as (electro-)chemical interactions of the drug molecules with the porous \ac{PMP} matrix and the medium.
In fact, it has been observed that the porous inner structure of the \acp{PMP} depends on $\Ri$ \cite{siepmann2005HowAutocatalysisAccelerates}.
Since $\Di$ models the effective mobility of the drug molecules inside this structure, the two model parameters $\Ri$ and $\Di$ are, hence, correlated.
To incorporate this effect into \eqref{eq:def:n_t}, an exponential model of the following form has been proposed in \cite{siepmann2005HowAutocatalysisAccelerates}
\begin{equation}
    \Di = \Dihat R^{\omega},\label{eq:def:Di}
\end{equation}
where $\Dihat$ in \si{\milli\meter\squared\per\hour} and $R = \Ri/(\SI{1}{\milli\meter})$ denote a scaling constant independent of the particle radius and the dimensionless normalized particle radius, respectively, and $\omega \geq 0$ indicates the degree to which $\Ri$ and $\Di$ are correlated.

Substituting \eqref{eq:def:Di} in \eqref{eq:def:n_t}, we obtain
\begin{align}
    x(t) &= \frac{6}{\pi^2} \sum_{n=1}^{\infty} \frac{\Ditilde
     \alpha_n^2}{n^2} \exp\left(-\Ditilde \alpha_n^2 t \right),\label{eq:n_t:omega}
\end{align}

where $\Ditilde = \Dihat / \SI{1}{\milli\meter\squared}$ and $\alpha_n = n \pi / R^{1 - \omega/2}$.

The value of $\omega$ is highly drug-specific, depending for example on the electrostatic interactions between the specific drug molecules and the \ac{PMP} polymers \cite{klose2008PlgaBasedDrug}.
However, since $\omega > 2$ would correspond to a positive correlation between $x(t)$ and $R$ in \eqref{eq:n_t:omega}, contradicting experimental observations \cite{klose2008PlgaBasedDrug}, we restrict our considerations to $\omega \in [0, 2)$.

From \eqref{eq:n_t:omega}, we observe that $x(t)$ is relatively sensitive with respect to changes in $R$ for values of $\omega$ close to $0$ and relatively insensitive for values of $\omega$ close to $2$.
Hence, the value of $\omega$ in \eqref{eq:def:Di} determines whether $R$ can be a suitable design parameter for the considered \ac{CDD} system.
In Section~\ref{sec:results}, we study the impact of $\omega$ on the sensitivity of the end-to-end drug release to changes in $R$ in a quantitative manner.

\subsubsection{Channel}

According to Section~\ref{sec:system_model:exp}, the drug molecules released from the \acp{PMP} propagate inside the \ac{BNC} fleece towards the release medium.
Hence, in the \ac{MC} framework, the \ac{BNC} fleece corresponds to the channel.

Since the \ac{BNC} fleece essentially behaves like a fluid medium (the water content is $>90\%$ \cite{iguchi2000BacterialCelluloseMasterpiece}), it is reasonable to assume that the \ac{CBD} propagation exhibits diffusion dynamics.
In line with this assumption, diffusion-controlled drug release from \ac{BNC} has also been reported in \cite{tangsatianpan2020ReleaseKineticModel}.
Now, since (i) drug molecules can leave the \ac{BNC} fleece only at its interface with the release medium, (ii) \acp{PMP} are uniformly distributed inside the \ac{BNC} patch, and (iii) the diffusive resistance at the interface layer is negligible as compared to the diffusive resistance inside the fleece \cite{bellmann2023SituFormationPolymer}, the channel impulse response, denoted by $h(t)$, is obtained by solving the one-dimensional diffusion equation for a closed interval with one Neumann (upper boundary of the fleece in Fig.~\ref{fig:system_model}) and one Dirichlet boundary (interface of \ac{BNC} fleece with release medium) and uniform initial condition as follows \cite[Eq.~(4.18)]{crank79}
\begin{equation}
    h(t) = \frac{8}{\pi^2} \sum_{m=1}^{\infty} \frac{\Do \beta_m}{(2m - 1)^2} \exp\left(-\Do \beta_m^2 t\right), \qquad t \geq 0, \label{eq:def:h_t}
\end{equation}
where $\Do$ denotes the diffusion coefficient of the \ac{CBD} inside the \ac{BNC} fleece, but outside the \acp{PMP}, in \si{\milli\meter\squared\per\hour} and $\beta_m = (2m -1) \pi/(2 a)$, with $a$ denoting the height of the \ac{BNC} fleece in \si{\milli\meter}.

Eq.~\eqref{eq:def:h_t} has been previously used to describe the drug release from a drug carrier film, e.g., in \cite{klose2008PlgaBasedDrug}.
For the system considered in this paper, we verify in Section~\ref{sec:results} with experimental data that \eqref{eq:def:h_t} is indeed a suitable channel model.

\subsubsection{Receiver}

The physical \ac{Rx} in the considered \ac{CDD} system is the release medium, cf.~Section~\ref{sec:system_model:exp}.
Conceptually, the \ac{Rx} is a transparent, molecule counting \ac{Rx}, i.e., it counts the cumulative amount of \ac{CBD} released into the release medium by time $t$ and does otherwise not interact with the drug molecules.
Since the considered system can be assumed to be linear and time-invariant, cf.~Section~\ref{sec:system_model:exp}, the {\em expected received signal}, $r(t)$, is obtained as $r(t) = \int_{0}^{t} x(\tau) * h(\tau) \dtau$, where $*$ denotes the convolution operator.
Substituting \eqref{eq:def:n_t} and \eqref{eq:def:h_t}, we obtain after straightforward integration
\begin{align}
    r&(t) =\nonumber\\
    &1 - \frac{6}{\pi^2} \sum_{n=1}^{\infty} \left[\frac{1}{n^2} \exp\left(-\frac
    {\Ditilde n^2 \pi^2}{R^{2-\omega}} t \right) - \sum_{m=1}^{\infty}\frac{1}{(2m - 1)^2} \right.\nonumber\\
    & \left. \frac{8 \Ditilde}{R^{2-\omega}} \frac{\exp\left( -\Ditilde n^2 \pi^2 t / R^{2-\omega} \right) - \exp\left(-\Do \beta_m^2 t\right)}{\Do \beta_m^2 - \Ditilde n^2 \pi^2/R^{2-\omega}}\right].\label{eq:r_t_full}
\end{align}

\section{Statistical Analysis}
\label{sec:analysis}
In this section, we use statistical analysis to study the impact of the \ac{PMP} radius $R$ on $r(t;R)$, where we write $r(t;R)$ instead of $r(t)$ to emphasize the dependence of $r(t)$ on $R$.
In this way, we address two important sources of uncertainty regarding $R$ that can arise in practice; (i) the variability of $R$ across different \acp{PMP} within the same \ac{BNC} fleece, (ii) variations in $R$ across different \ac{BNC} fleeces. 
Since $R$ is one of the potential design parameters, the results presented in this section constitute a step towards understanding the possibilities and limits of tuning the considered \ac{CDD} system by adjusting the \ac{PMP} size.

\subsection{PMP Size Distribution in one BNC fleece}\label{sec:analysis:radius_dist}

In practice, the formation of the \acp{PMP} is a random process and, hence, not all \acp{PMP} in one \ac{BNC} fleece have the same radius, i.e., $R$ can be different for every \ac{PMP}.
Furthermore, the amount of drug molecules encapsulated in a \ac{PMP} during its synthesis may depend on the \ac{PMP}'s size.
However, to the best of our knowledge, no data in this regard is available.
Hence, in this paper, we consider two different hypotheses; namely, that the amount of drug molecules enclosed in a \ac{PMP} is (i) {\em independent} of the \ac{PMP}'s size, (ii) {\em proportional to the volume} of the \ac{PMP}.
The drug release under Hypothesis (i) is studied analytically in this section, while (ii) is evaluated using Monte-Carlo simulations.
Numerical results presented in Section~\ref{sec:results} show that for realistic \ac{PMP} size distributions, the drug release under both hypotheses is almost identical.

To account for different \ac{PMP} sizes, $R$ is in the following modeled as a continuous, non-negative random variable with \ac{pdf} $f_R(x)$.
Under Hypothesis (i) the total drug release into the release medium in the presence of different sized \acp{PMP} can be expressed as $\Exp{r(t;R)}$, where $\Exp{\cdot}$ denotes the expectation operator with respect to $f_R(x)$.
Furthermore, since $r(t)$ is a nonlinear function of $R$, in general,
\begin{equation}
    \mu_r(t) = \Exp{r(t;R)} \neq r(t;\Exp{R}).\label{eq:Ert_vs_rER}
\end{equation}

The true distribution of $R$ is unknown and depends on the stochastic synthesis process of the \acp{PMP}.
However, for a specific realization of this process, empirical data on the true distribution of $R$ is available \cite{bellmann2023SituFormationPolymer}.
In order to obtain an {\em analytical model} for the distribution of $R$, we start from the hypothesis that the rate parameters of the exponential distributions in \eqref{eq:n_t:omega} are distributed according to the Gamma distribution with shape and rate parameters $\gamma_n$ and $\zeta_n$, respectively, i.e.,
\begin{equation}
\Ditilde \alpha_n^2 \sim \mathrm{Gamma}\left(\gamma_n, \zeta_n\right).\label{eq:gamma}
\end{equation}
The Gamma distribution is a commonly used prior for the rate parameter of the exponential distribution \cite{gelman2013BayesianDataAnalysis}.
In order to match \eqref{eq:gamma} to the moments of $R$ that can be experimentally observed, we establish the following theorem.

\begin{theorem}\label{thm:moments_R}
    Let $\Ditilde \alpha_n^2 \sim \mathrm{Gamma}\left(\gamma_n, \zeta_n\right)$. Then,
    \begin{align}
        \Exp{R} = \mu_R &= \frac{\Gamma(\gamma - 1/(2-\omega))}{\Gamma(\gamma)}\zeta^{1/(2-\omega)},\label{eq:exp_R}\\
        \Exp{R^2} = \sigma_R^2 + \mu_R^2 &= \frac{\Gamma(\gamma - 2/(2-\omega))}{\Gamma(\gamma)}\zeta^{2/(2-\omega)},\label{eq:exp_R2}
    \end{align}
    where $\zeta = \Di n^2 \pi^2 \zeta_n$ and $\gamma = \gamma_n$ are independent of $n$, and $\sigma_R^2$ and $\Gamma(\cdot)$ denote the variance of $R$ and  the Gamma function, respectively.
\end{theorem}
\begin{IEEEproof}
    \ifx \arXiv \undefined
    It follows from the scaling property of the Gamma distribution that $R^{-(2 - \omega)} \sim \mathrm{Gamma}\left(\gamma_n, \Di n^2 \pi^2 \zeta_n\right)$. Since this holds for any $n$, $R^{-(2 - \omega)} \sim \mathrm{Gamma}\left(\gamma, \zeta\right)$ follows.
    The moments of $R$ are then computed by applying the transformation theorem for \acp{RV}.
    For the complete proof, please see the arxiv version of this paper \cite{lotter2023ARXIVMicroparticleBasedControlled}.
    \else
    Please see Appendix~\ref{sec:app:thm_1}.
    \fi
\end{IEEEproof}

The following corollary of Theorem~\ref{thm:moments_R} allows us to choose $\gamma_n$ and $\zeta_n$ in \eqref{eq:gamma} according to the experimentally observed statistics of $R$.
\begin{corollary}\label{cor:params_gamma}
    From the first and second central moments of $R$, $\mu_R$ and $\sigma_R^2$, $\gamma$ and $\zeta$ are obtained from \eqref{eq:exp_R} and \eqref{eq:exp_R2} as the solutions to the following system of equations:
    \begin{align}
        \left[\mu_R \frac{\Gamma(\gamma)}{\Gamma(\gamma - 1/(2-\omega))}\right]^{(2-\omega)} - \zeta &= 0,\\
        \Gamma(\gamma)\frac{\Gamma(\gamma - 2/(2-\omega))}{\Gamma\left(\gamma - 1/(2-\omega)\right)^2} - \left(1 + \frac{\sigma_R^2}{\mu_R^2}\right) &= 0.
    \end{align}
\end{corollary}

We confirm that the statistical model for the \ac{PMP} sizes proposed in this paper is indeed accurate by comparing it to the experimentally observed particle size distribution from \cite{bellmann2023SituFormationPolymer}.
To this end, we set $\mu_R=\SI{1}{\micro\meter}$ and $\sigma_R=\SI{0.12}{\micro\meter}$ in agreement with \cite{bellmann2023SituFormationPolymer} (data set EL10).
To assess the quality of the fit, $10^8$ random samples are drawn according to \eqref{eq:gamma} and, as reference, from a Gaussian distribution, $\mathcal{N}(\mu_R, \sigma^2_R)$, with mean $\mu_R$ and variance $\sigma_R^2$, respectively.
Table~\ref{tab:fit_pmp_size} shows the agreement of both models with the experimental data in terms of their respective Kullback-Leibler divergences, where the Kullback-Leibler divergence between probability mass vectors $p = [p_1,\ldots,p_N]$ and $q=[q_1,\ldots,q_N]$ is defined as $K_{\mathrm{LD}}(p||q) = \sum_{i=1}^{N} p_i \log (p_i/q_i)$, where $\log(\cdot)$ denotes the natural logarithm.
Here, $p$ and $q$ correspond to the normalized empirical frequencies of the data and the random samples, respectively, after binning both in bins of width \SI{0.14}{\micro\meter} as in \cite{bellmann2023SituFormationPolymer}.
As can be observed from Table~\ref{tab:fit_pmp_size}, both considered models are accurate, but \eqref{eq:gamma} provides a even slightly better fit of the \ac{PMP} size distribution than the Gaussian benchmark model.

\begin{table}
    \vspace*{0.07in}
    \centering
    \caption{Parameter Size Distribution}
    \footnotesize
    \begin{tabular}{|c||c|c|}\hline
        $K_{\mathrm{LD}}(p||q)$ & This paper \eqref{eq:gamma} & $\mathcal{N}(\mu_R, \sigma_R)$\\\hline\hline
        Experiment \cite{bellmann2023SituFormationPolymer} & $2.80 \cdot 10^{-2}$ & $2.85 \cdot 10^{-2}$\\\hline
    \end{tabular}
    \label{tab:fit_pmp_size}
    \vspace*{-4mm}
\end{table}

Now that the analytical model for $R$ in \eqref{eq:gamma} is confirmed, $\Exp{r(t;R)}$ can be characterized as follows.

\begin{theorem}\label{thm:mu_r}
    Let $\Ditilde \alpha_n^2 \sim \mathrm{Gamma}\left(\gamma_n, \zeta_n\right)$. Then,
    \begin{align}
        &\mu_r(t) = 1 - \frac{6}{\pi^2} \sum_{n = 1}^{\infty} \frac{1}{n^2} \left[ \left(\frac{\zeta}{\zeta + \Ditilde n^2 \pi^2 t}\right)^{\gamma}\right.\nonumber\\
        &\left. + 8 \sum_{m=1}^{\infty} \frac{\gamma \zeta^{\gamma} \Ditilde n^2}{\left(2 m - 1\right)^2}  \bigintsss_{0}^{t} \frac{\exp\left(-\Do \beta_m^2\left(t-\tau\right)\right)}{\left(\zeta + \Ditilde n^2 \pi^2 \tau\right)^{(\gamma + 1)}}\,\dtau \right].
    \end{align}
\end{theorem}
\begin{IEEEproof}
    \ifx \arXiv \undefined
    The theorem can be proved by substituting the \ac{pdf} of the Gamma distribution and computing the expectations of the single terms in \eqref{eq:r_t_full} after writing the terms of the last sum as convolutions.
    For the complete proof, please see the arxiv version of this paper \cite{lotter2023ARXIVMicroparticleBasedControlled}.
    \else
    Please see Appendix~\ref{sec:app:thm_2}.
    \fi
\end{IEEEproof}

Theorem~\ref{thm:mu_r} is utilized in Section~\ref{sec:results} to show that for realistic values of $\mu_R$ and $\sigma_R$, $\mu_r(t) \approx r(t;\mu_R)$, i.e., that \acp{PMP} of the same size can be assumed if only one single \ac{BNC} fleece is considered.

\subsection{PMP Size Variations Across Different BNC Fleeces}

Exploiting the results obtained in the previous section, we will now focus on the variability of $r(t;R)$ due to different \ac{PMP} synthesis conditions in different \ac{BNC} fleeces.
To this end, we assume that the \acp{PMP} in each fleece have equal size, but the sizes of the \acp{PMP} in different fleeces vary from fleece to fleece.
Hence, we consider $R$ an unknown parameter for which only statistical knowledge is available.
Specifically, we are interested in computing the variance of $r(t;R)$, $\sigma_r^2(t)$.
Similar as in the previous section, we assume that the rate parameters of the exponential functions in \eqref{eq:n_t:omega} are distributed according to \eqref{eq:gamma} and obtain the following result.

\begin{theorem}\label{thm:e_r2}
    Let $\Ditilde \alpha_n^2 \sim \mathrm{Gamma}\left(\gamma_n, \zeta_n\right)$. Then,
    \begin{align}
        &\sigma_r^2(t) = \sum_{n,l=1}^{\infty} a_n a_l \mathcal{A}_{n,l}(t) + 2 \sum_{n,m,l=1}^{\infty} b_{n,m,l} \mathcal{B}_{n,m,l}(t)\nonumber\\
        &\qquad + \sum_{n,m,l,k=1}^{\infty} c_{n,m,l,k} \mathcal{C}_{n,m,l,k}(t) - \left[1 - \mu_r(t)\right]^2,
    \end{align}
    where $a_n = 6/\left(\pi^2 n^2\right)$, $b_{n,m,l} = 8 a_n a_l/\left[\pi^2 \left(2 m - 1\right)^2\right]$, $c_{n,m,l,k} = b_{n,m,l} b_{n,k,l}/\left(a_n a_l\right)$,%
    \begin{align}
        \mathcal{A}_{n,l}(t) &= \left[\frac{\zeta}{\zeta + \Ditilde \pi^2 \left(n^2 + l^2\right) t}\right]^{\gamma},\\
        \mathcal{B}_{n,m,l}(t) &= \gamma \zeta^{\gamma} \Ditilde n^2 \pi^2 \,\,\int\limits_0^t \exp\left(-\Do \beta_m^2 (t-\tau)\right) / \nonumber\\
        &\quad \left(\zeta + \Ditilde \pi^2 (n^2 \tau + l^2 t)\right)^{(\gamma+1)} \dtau,\\
        \mathcal{C}_{n,m,l,k}(t) &= \gamma (\gamma + 1) \zeta^{\gamma} \Ditilde^2 n^2 l^2 \pi^4 \,\,\int\limits_0^t \int\limits_0^{\theta} \mathcal{D}_{n,m,l,k}(t, \tau, \theta) \nonumber\\
        &\quad + \mathcal{D}_{l,k,n,m}(t, \tau, \theta) \dtau \dtheta,
    \end{align}
    and 
    \begin{multline}
    \mathcal{D}_{n,m,l,k}(t, \tau, \theta) = \exp\left(-\Do (\beta_m^2 (\theta-\tau) + \beta_k^2 \theta)\right)/\\
    \quad\left(\zeta + \Ditilde \pi^2 (n^2 (t + \tau - \theta) + l^2 (t - \theta))\right)^{(\gamma+2)}.
    \end{multline}
\end{theorem}
\begin{IEEEproof}
    \ifx \arXiv \undefined
    The expectations of the cross-terms in $r^2(t)$ can be computed after simplifying the terms in the Fourier domain.
    For the complete proof, please see the arxiv version of this paper \cite{lotter2023ARXIVMicroparticleBasedControlled}.
    \else
    Please see Appendix~\ref{sec:app:thm_3}.
    \fi
\end{IEEEproof}

Although the computation of $\sigma_r^2(t)$ involves the numerical evaluation of integrals, these integrals are definite integrals over finite intervals and can be computed efficiently using Gaussian quadrature.

\section{Numerical Results}
\label{sec:results}
In this section, numerical results for the analysis conducted in the previous section are presented and validated with experimental data and stochastic simulations. Unless otherwise noted, the default parameter values listed in Table~\ref{tab:parameter_values} are used in agreement with experimental data in \cite{bellmann2023SituFormationPolymer}. The free parameters $\Di$ and $\Do$ are set via exhaustive search to minimize the least-squares error between the analytical model and the experimental data.
As a goodness-of-fit measure, we report the \ac{MSE} between model and data instead of the commonly used $R^2$ value, since the $R^2$ value is not meaningful for nonlinear regression as considered in this paper \cite{spiess2010EvaluationR2As}.

\begin{table}
    \vspace*{0.07in}
    \centering
    \caption{Default Parameter Values \cite{bellmann2023SituFormationPolymer}}
    \footnotesize
    \begin{tabular}{|c|c|c|c|}\hline
        \textbf{Parameter} & \textbf{Default Value} & \textbf{Parameter} & \textbf{Default Value}\\\hline
        $a$ & \SI{3.54}{\milli\meter} & $\mu_R$ & \SI{1}{\micro\meter}\\\hline
        $\omega$ & $0$ & $\sigma_R$ & \SI{0.12}{\micro\meter}\\\hline
    \end{tabular}
    \label{tab:parameter_values}
    \vspace*{-4mm}
\end{table}

\subsection{Drug Release from One BNC Fleece}

\begin{figure}
    \centering
    \includegraphics[width=.4\textwidth]{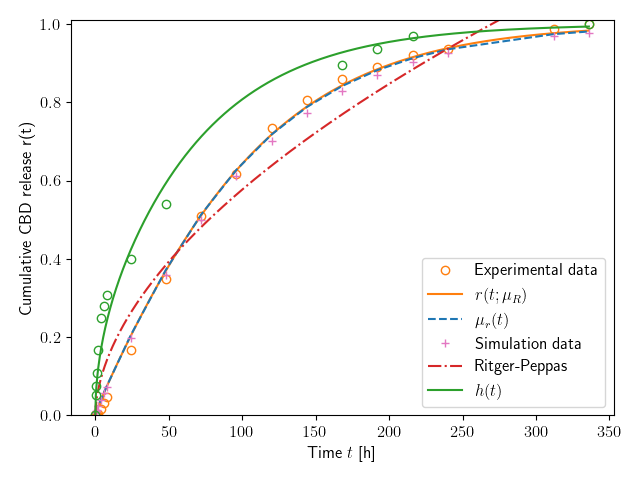}
    \vspace*{-3mm}
    \caption{\ac{CBD} release from \ac{BNC} fleece with (orange) and without (green) \acp{PMP}. Circle and cross markers denote experimental data from \cite{bellmann2023SituFormationPolymer} (orange: data set EL10, green: data set propylene glycol) and simulation results, respectively.}
    \label{fig:cbd_release}
    \vspace*{-6mm}
\end{figure}

First, we exploit data from experiments in which the \ac{CBD} molecules are {\em not} embedded into \acp{PMP}, but released from a propylene glycol solution in the \ac{BNC} fleece \cite{bellmann2023SituFormationPolymer} to verify the proposed channel model \eqref{eq:def:h_t} in isolation.
In the considered scenario, the \ac{CBD} molecules are released instantaneously in the \ac{BNC} fleece, hence, the measured \ac{CBD} concentration at the \ac{Rx} corresponds to the channel impulse response in \eqref{eq:def:h_t}.
Fig.~\ref{fig:cbd_release} shows $h(t)$ for $\Do = 7.3 \cdot 10^{-2} \,\si{\milli\meter\squared\per\hour}$ along with the experimental data and we observe that the agreement is excellent ($\textrm{MSE} = 1.6 \cdot 10^{-3}$).

Next, we consider the regular case, when the \ac{CBD} molecules are released from \acp{PMP} inside the \ac{BNC} fleece.
Fig.~\ref{fig:cbd_release} shows experimental data for this case along with $\mu_r(t)$, $r(t;\mu_R)$, and results from Monte-Carlos simulations, where, in the analytical and simulation results, $\Di=1.62 \cdot 10^{-9} \,\si{\milli\meter\squared\per\hour}$ and $\Do=8.13 \cdot 10^{-2} \,\si{\milli\meter\squared\per\hour}$.
The Monte-Carlo simulations are performed by drawing $10^4$ random samples according to \eqref{eq:gamma} and then averaging \eqref{eq:r_t_full} over these samples.
In the simulations, it is assumed that the amount of \ac{CBD} released from each \ac{PMP} is proportional to its volume.
We observe from Fig.~\ref{fig:cbd_release} that the agreement between $r(t;\mu_R)$ and the experimental data is excellent ($\textrm{MSE}=3.1 \cdot 10^{-4}$).
Furthermore, we observe that the deviation between $\mu_r(t)$, $r(t;\mu_R)$, and the Monte-Carlo simulations is negligible.
This latter observation has two important consequences.
First, the agreement between Monte-Carlo simulations and $\mu_r(t)$ confirms that, for the considered \ac{PMP} radius distribution, the assumption that all \acp{PMP} release the same total amount of \ac{CBD} needed for our analysis is not critical, cf.~Section~\ref{sec:analysis:radius_dist}.
Second, for the given \ac{PMP} size statistics, since $\mu_r(t) \approx r(t;\mu_R)$, we can safely assume that all \acp{PMP} inside the same \ac{BNC} fleece have the same size.
Fig.~\ref{fig:cbd_release} shows also that the proposed model significantly outperforms the Ritger-Peppas model \eqref{eq:ritger-peppas}, which yields $\textrm{MSE}= 5.2 \cdot 10^{-3}$, although both models involve two fitting parameters.

\subsection{Unknown PMP Sizes Across Different BNC Fleeces}\label{sec:results:statistical_analysis}

Finally, we study the impact of the correlation between $R$ and $\Di$ as expressed in \eqref{eq:def:Di} on $r(t;R)$ assuming that $R$ is fixed, but we do not know its true value.
To this end, we exploit Theorems~\ref{thm:mu_r} and \ref{thm:e_r2} and compute $\sigma_r(t)$ as a measure of the uncertainty about $r(t;R)$ that results from knowing $R$ only statistically.
In practice, $\sigma_r(t)$ corresponds to the intrinsic variability in the drug releases from different \ac{BNC} fleeces and is therefore an important parameter to consider for system design.
To reflect the higher variability of \ac{PMP} radii across different \ac{BNC} fleeces as compared to the \ac{PMP} size variation within one fleece, we assume a slightly larger $\sigma_R = 0.24 \,\si{\micro\meter}$ here than in the previous section.

\begin{figure}[t]
    \centering
    \includegraphics[width=.4\textwidth]{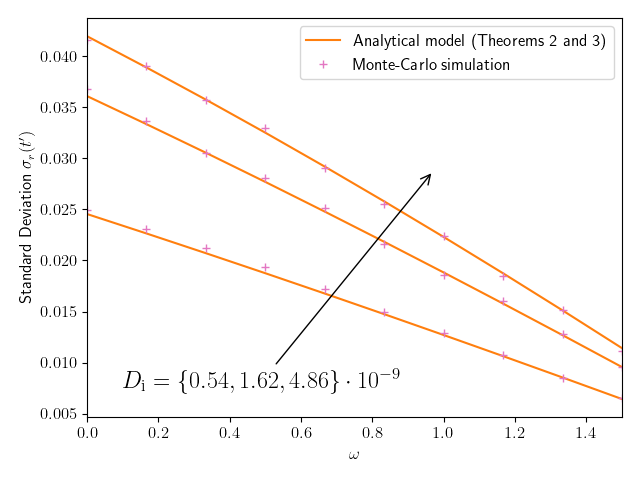}
    \vspace*{-3mm}
    \caption{$\sigma_r(t')$ as a function of $\omega$ at $t'=24$ h as obtained from Theorems~\ref{thm:mu_r} and \ref{thm:e_r2} (lines) and from Monte-Carlo simulations (markers), respectively, for different values of $\Di$.}
    \label{fig:R_unknown}
    \vspace*{-6mm}
\end{figure}

Fig.~\ref{fig:R_unknown} shows $\sigma_r(t')$ as a function of $\omega$ at time $t'=24$ h for different values of $\Di$, i.e., for each $\omega$, $\Dihat$ in \eqref{eq:def:Di} is adjusted such that $\Di$ is the same for all values of $\omega$.
In this way, $r(t')$ is the same for all considered values of $\omega$.
We observe from Fig.~\ref{fig:R_unknown} that $\omega$ does indeed have a significant impact on the variability of the drug release.
Specifically, $\sigma_r(t')$ decreases as $\omega$ increases.
Hence, the stronger $R$ and $\Di$ are correlated, the more deterministic is $r(t')$.
Furthermore, we observe from Fig.~\ref{fig:R_unknown} that this effect is more pronounced for large values of $\Di$.
When $\omega \approx 0$, large values of $\Di$ incur large values of $\sigma_r(t')$, while $\sigma_r(t')$ is comparatively small for small values of $\Di$.
As $\omega$ increases, the difference between the $\sigma_r(t')$ for different values of $\Di$ is reduced.
Hence, when $R$ and $\Di$ are weakly correlated, the drug release is more sensitive towards variations in the \ac{PMP} radius when the drug mobility inside the \acp{PMP} is large, while for strongly correlated $R$ and $\Di$ the sensitivity of the drug release to variations in the \ac{PMP} radius is similar for different drug mobilities.
From this observation, we conclude that more control over the \ac{PMP} radius is required to ensure deterministic drug release for drug/\ac{PMP} constellations for which $\omega$ is small as compared to ones for which $\omega$ is large.

\vspace*{-1mm}
\section{Conclusion}
\label{sec:conclusion}
\vspace*{-1mm}
In this paper, we have proposed a novel model for microparticle-based \ac{CDD} from \ac{BNC}.
The proposed model has been validated with experimental data and shows excellent agreement while keeping the dimension of the parameter space low.
Furthermore, a comprehensive statistical analysis has been presented that allows the characterization of the impact of the microparticle radius on the drug release.
The presented results reveal that the sensitivity of the drug release with respect to changes of the particle radius depends on the degree to which particle radius and particle morphology are correlated.

To gain further insights on which specific drug/microparticle combinations can support particular drug release profiles, the values of $\omega$ corresponding to these combinations need to be characterized in future work.
Such a characterization together with the results presented in this paper will guide the development of more and more refined microparticle-based \ac{CDD} systems for specific applications that require high-precision control and reproducibility of drug release profiles.

\ifx \arXiv \undefined
\else
\appendix
\section{}
\subsection{Proof of Theorem \ref{thm:moments_R}}\label{sec:app:thm_1}

To simplify the notation, let $\omtil = 2 - \omega$.
Then, since $\Ditilde \alpha_n^2 = \Ditilde n^2 \pi^2 / R^{\omtil} \sim \mathrm{Gamma}\left(\gamma_n, \zeta_n\right)$, it follows from the scaling property of the Gamma distribution that $R^{-\omtil} \sim \mathrm{Gamma}\left(\gamma_n, \Ditilde n^2 \pi^2 \zeta_n\right)$. Since this holds for any $n$, $R^{-\omtil} \sim \mathrm{Gamma}\left(\gamma, \zeta\right)$.
Furthermore, by definition of the inverse Gamma distribution, we have $R^{\omtil} \sim \mathrm{InvGamma}\left(\gamma, \zeta\right)$, where $\mathrm{InvGamma}\left(\alpha, \beta\right)$ denotes the inverse Gamma distribution with shape and rate parameters $\alpha$ and $\beta$, respectively. 
Now, according to the transformation theorem for \acp{RV}, the \ac{pdf} of $R$, $f_R(x)$, is obtained from the \ac{pdf} of $R^{\omtil}$, $f_{R^{\omtil}}(x)$, as follows
\begin{align}
    f_R(x) = f_{R^{\omtil}}(x^{\omtil}) \frac{\mathrm{d} x^{\omtil}}{\mathrm{d} x} = \frac{\omtil \zeta^\gamma}{\Gamma(\gamma)} x^{-(\omtil \gamma + 1)} \exp(-\frac{\zeta}{x^{\omtil}}),
\end{align}
where we have substituted the \ac{pdf} of the inverse Gamma distribution for $f_{R^{\omtil}}(x)$.

The following computations yield $\Exp{R}$ and $\Exp{R^2}$ as required by the theorem.
\begin{align}
    \Exp{R} &= \frac{\omtil \zeta^\gamma}{\Gamma(\gamma)} \int\limits_{0}^{\infty} x^{-\omtil \gamma} \exp\left(-\frac{\zeta}{x^{\omtil}}\right) \,\mathrm{d}x\nonumber\\
    &= \frac{\zeta^\gamma}{\Gamma(\gamma)} \int\limits_{0}^{\infty} u^{-\gamma} \exp\left(-\frac{\zeta}{u}\right) u^{(1/\omtil) - 1} \,\mathrm{d}u\nonumber\\
    &= \frac{\zeta^\gamma}{\Gamma(\gamma)} \int\limits_{0}^{\infty} u^{-[\gamma - (1/\omtil)+1]} \exp\left(-\frac{\zeta}{u}\right) \,\mathrm{d}u\nonumber\\
    &= \frac{\Gamma(\gamma-1/\omtil)}{\Gamma(\gamma)}\frac{\zeta^\gamma}{\zeta^{\gamma - 1/\omtil}} = \frac{\Gamma(\gamma-1/\omtil)}{\Gamma(\gamma)}\zeta^{1/\omtil},\label{eq:app:thm_1:r1}
\end{align}
\begin{align}
    \Exp{R^2} &= \frac{\omtil \zeta^\gamma}{\Gamma(\gamma)} \int\limits_{0}^{\infty} x^{-\omtil \gamma + 1} \exp\left(-\frac{\zeta}{x^{\omtil}}\right) \,\mathrm{d}x\nonumber\\
    &= \frac{\zeta^\gamma}{\Gamma(\gamma)} \int\limits_{0}^{\infty} u^{-\gamma + 1/\omtil} \exp\left(-\frac{\zeta}{u}\right) u^{(1/\omtil) - 1} \,\mathrm{d}u\nonumber\\
    &= \frac{\zeta^\gamma}{\Gamma(\gamma)} \int\limits_{0}^{\infty} u^{-[\gamma - (2/\omtil)+1]} \exp\left(-\frac{\zeta}{u}\right) \,\mathrm{d}u\nonumber\\
    &= \frac{\Gamma(\gamma-2/\omtil)}{\Gamma(\gamma)}\frac{\zeta^\gamma}{\zeta^{\gamma - 2/\omtil}} = \frac{\Gamma(\gamma-2/\omtil)}{\Gamma(\gamma)}\zeta^{2/\omtil},\label{eq:app:thm_1:r2}
\end{align}
where we used the substitution $u = x^{\omtil}$ in both equations.
Substituting the definition of $\omtil$ into \eqref{eq:app:thm_1:r1}, \eqref{eq:app:thm_1:r2}, Theorem~\ref{thm:moments_R} follows. This completes the proof.

\subsection{Proof of Theorem \ref{thm:mu_r}}\label{sec:app:thm_2}

In order to compute $\Exp{r(t;R)}$, we exploit $\Ditilde \alpha_n^2 \sim \mathrm{Gamma}\left(\gamma, \zeta_n\right)$, cf.~Theorem~\ref{thm:moments_R}, and compute the expectations of the summands in \eqref{eq:r_t_full} separately.
We obtain
\begin{align}
    &\Exp{\exp\left(- \Ditilde \alpha_n^2 t \right)}\nonumber\\ 
    &\quad= \frac{\zeta_n^\gamma}{\Gamma(\gamma)} \int\limits_{0}^{\infty} \exp(- t x) x^{\gamma-1} \exp(-\zeta_n x) \dx \nonumber\\
    &\quad=\frac{\zeta_n^\gamma}{\Gamma(\gamma)} \int\limits_{0}^{\infty} x^{\gamma-1} \exp(-(\zeta_n + t) x) \dx \nonumber\\
    &\quad=\frac{\zeta_n^\gamma}{(\zeta_n + 1)^\gamma} = \left(\frac{\zeta}{\zeta + \Ditilde n^2 \pi^2 t}\right)^{\gamma},
\end{align}
and,
\begin{align}
    &\Exp{\Ditilde \alpha_n^2 \frac{\exp\left( -\Ditilde \alpha_n^2 t\right) - \exp\left(-\Do \beta_m^2 t\right)}{\Do \beta_m^2 - \Ditilde \alpha_n^2}}\nonumber\\
    &\,=\frac{\zeta_n^\gamma}{\Gamma(\gamma)} \int\limits_{0}^{\infty} \frac{\exp\left( -x t\right) - \exp\left(-\Do \beta_m^2 t\right)}{\Do \beta_m^2 - x} x^{\gamma} \exp(-\zeta_n x) \dx\nonumber\\
    &\,=\frac{\zeta_n^\gamma}{\Gamma(\gamma)} \int\limits_{0}^{\infty} \frac{\exp\left( -(\zeta_n + t) x\right) - \exp\left(-\Do \beta_m^2 t - \zeta_n x\right)}{\Do \beta_m^2 - x} x^{\gamma} \dx\nonumber\\
    &\,\stackrel{(a)}{=}\frac{\zeta_n^\gamma}{\Gamma(\gamma)} \dint{0}{\infty} \dint{0}{t} \exp\left(-(\tau + \zeta_n) x - \Do \beta_m^2 (t - \tau)\right) x^\gamma \dtau \dx \nonumber\\
    &\,=\dint{0}{t} \frac{\zeta_n^{\gamma} \gamma}{\left(\zeta_n + \tau\right)^{(\gamma + 1)}} \exp(-\Do \beta_m^2 (t - \tau)) \dtau,
\end{align}
where we have utilized the fact that the fraction in the integrand on the left-hand side of $(a)$ can be written in terms of a convolution in $t$ and that the integrals on the right-hand side of $(a)$ can be interchanged.
Substituting the definition of $\zeta_n$ from Theorem~\ref{thm:moments_R} and collecting terms, Theorem~\ref{thm:mu_r} follows.
This completes the proof.

\subsection{Proof of Theorem \ref{thm:e_r2}}\label{sec:app:thm_3}

First, let us rewrite $r(t)$ as follows
\begin{align}
    r(t) = 1 - \sum_{n=1}^{\infty} a_n A_n(t) + \sum_{n,m=1}^{\infty} g_{n,m} G_{n,m}(t),
\end{align}
where $g_{n,m} = a_n 8 \pi^2 / \left(2m - 1\right)^2$, $A_n(t) = \exp(-\Ditilde \alpha_n^2) t$, and
\begin{equation}
    G_{n,m}(t) = \Ditilde \alpha_n^2 \frac{\exp(-\Ditilde \alpha_n^2 t) - \exp(-\Do \beta_m^2 t)}{\Do \beta_m^2 - \Ditilde \alpha_n^2}. 
\end{equation}
Then,
\begin{align}
    \sigma_r^2(t) &= \Exp{r(t)^2} - \mu_r(t)^2\nonumber\\
    &= \Exp{\left[\sum_{n=1}^{\infty} a_n A_n(t) + \sum_{n,m=1}^{\infty} g_{n,m} G_{n,m}(t)\right]^2} + 1\nonumber\\
    &{}\quad- 2 \Exp{\sum_{n=1}^{\infty} a_n A_n(t) + \sum_{n,m=1}^{\infty} g_{n,m} G_{n,m}(t)} - \mu_r(t)^2\nonumber\\
    &= \Exp{\left[\sum_{n=1}^{\infty} a_n A_n(t) + \sum_{n,m=1}^{\infty} g_{n,m} G_{n,m}(t)\right]^2} + 1 \nonumber\\
    &{}\quad - 2 \left[1 - \mu_r(t)\right] - \mu_r(t)^2 \nonumber\\
    &= \Exp{\left[\sum_{n=1}^{\infty} a_n A_n(t) + \sum_{n,m=1}^{\infty} g_{n,m} G_{n,m}(t)\right]^2} \nonumber\\
    &{}\quad - \left[1 - \mu_r(t)\right]^2.\label{eq:app:thm_3:cross-terms}
\end{align}

Hence, in order to compute $\sigma_r(t)^2$, after opening the square in the expectation on the right-hand side of \eqref{eq:app:thm_3:cross-terms}, the following terms need to be computed:
\begin{align}
    &\Exp{A_n(t) A_l(t)},\label{eq:app:thm_3:list:A_A}\\
    &\Exp{A_n(t) G_{l,k}(t)},\label{eq:app:thm_3:list:A_G}\\
    &\Exp{G_{n,m}(t) G_{l,k}(t)},\label{eq:app:thm_3:list:G_G}
\end{align}
where $l,k$ are positive integers.
From the proof of Theorem~\ref{thm:moments_R}, we recall that $R^{-(2-\omega))} \sim \mathrm{Gamma}\left(\gamma, \zeta\right)$.
Hence, substituting the \ac{pdf} of the Gamma distribution in \eqref{eq:app:thm_3:list:A_A}, we obtain
\begin{align}
    &\Exp{A_n(t) A_l(t)} \nonumber\\
    &\quad=\frac{\zeta^\gamma}{\Gamma(\gamma)} \dint{0}{\infty} \exp\left(-\Ditilde (n^2 + l^2) \pi^2 t x\right) x^{\gamma-1} \exp(-\zeta x) \dx \nonumber\\
    &\quad=\left[\frac{\zeta}{\zeta + \Ditilde \pi^2 (n^2 + l^2)t}\right]^{\gamma}.\label{eq:app:thm_3:A_n_A_l}
\end{align}

Before we proceed to computing \eqref{eq:app:thm_3:list:A_G} and \eqref{eq:app:thm_3:list:G_G}, we require some more notation.
In particular, for any square-integrable function $f(t)$, let its {\em Fourier transform} $\hat{f}(\xi)$ be defined as
\begin{align}
    \hat{f}(\xi) = \mathbb{F}\left[f(t)\right] = \dint{-\infty}{\infty} f(t) \exp(-j \xi t) \,\dt,
\end{align}
where $j$ denotes the imaginary unit.
Similarly, let the {\em inverse Fourier transform} of a function $\hat{f}(\xi)$ be defined as follows
\begin{align}
    f(t) = \mathbb{F}^{-1}\left[\hat{f}(\xi)\right] = \frac{1}{2 \pi} \dint{-\infty}{\infty} \hat{f}(\xi) \exp(j \xi t) \,\mathrm{d}\xi.
\end{align}
With these definitions, we obtain
\begin{align}
    &\mathbb{F}\left[A_n(t) G_{l,k}(t)\right] = \frac{\Ditilde \alpha_n^2}{\Do \beta_m^2 - \Ditilde \alpha_n^2} \nonumber\\
    &\quad \cdot \left[ \frac{1}{\Ditilde (\alpha_n^2 + \alpha_l^2) + j \xi} -  \frac{1}{\Do \beta_m^2 + \Ditilde \alpha_l^2 + j \xi}\right],\nonumber\\
    &\,= \frac{\Ditilde \alpha_n^2}{\left[\Ditilde(\alpha_n^2 + \alpha_l^2) + j \xi\right] \left(\Do \beta_m^2 + \Ditilde \alpha_l^2 + j \xi \right)},
\end{align}
where we have exploited repeatedly that $t \geq 0$ and $\mathbb{F}\left[\exp(-c t) u(t)\right] = 1/(c + j \xi)$, where $c>0$ is any positive constant and $u(t)$ denotes the unit step function.
Since multiplication in the Fourier domain corresponds to convolution in the time domain, we obtain
\begin{align}
    &\Exp{A_n(t) G_{l,k}(t)} = \Exp{\mathbb{F}^{-1}\left[\mathbb{F}\left[A_n(t) G_{l,k}(t)\right]\right]} \nonumber\\
    &\,= \mathbb{E}\left\lbrace\Ditilde \alpha_n^2 \dint{0}{t} \exp(-\Ditilde (\alpha_n^2 + \alpha_l^2) \tau)\right.\nonumber\\
    &\quad \left. \cdot\exp\left(-(\Do \beta_m^2 + \Ditilde \alpha_l^2)(t -\tau)\right) \dtau\right\rbrace \nonumber\\
    &\,= \mathbb{E}\left\lbrace\Ditilde n^2 \pi^2 R^{-(2-\omega)} \dint{0}{t} \exp(-\Do \beta_m^2 (t-\tau)) \right. \nonumber\\
    &\quad \left. \cdot\exp\left(-\Ditilde \pi^2 (n^2 \tau + l^2 t) R^{-(2-\omega)}\right) \dtau \right\rbrace.\label{eq:app:thm_3:A_n_G_l_k}
\end{align}
Substituting the \ac{pdf} of the Gamma distribution in \eqref{eq:app:thm_3:A_n_G_l_k}, we proceed by exchanging integrals as in the second part of the proof of Theorem~\ref{thm:mu_r} and obtain after some straightforward computations
\begin{align}
    &\Exp{A_n(t) G_{l,k}(t)}\nonumber\\
    &\quad = \gamma \Ditilde n^2 \pi^2 \zeta^{\gamma} \bigintsss_{0}^{t} \frac{\exp\left(-\Do \beta_m^2 (t - \tau)\right)}{\left[\zeta + \Ditilde \pi^2 (n^2 \tau + l^2 t)\right]^{\gamma + 1}} \dtau.\label{eq:app:thm_3:A_n_G_l_k_final}
\end{align}

The computation of \eqref{eq:app:thm_3:list:G_G} follows the same strategy.
After some simplification in the Fourier domain, $\mathbb{F}\left[G_{n,m}(t) G_{l,k}(t)\right]$ is obtained as shown in \eqref{eq:app:thm_3:G_G} on the next page.
Applying the inverse Fourier transform to \eqref{eq:app:thm_3:G_G}, exploiting again that multiplication in the Fourier domain corresponds to convolution in the time domain, and substituting the \ac{pdf} of the Gamma function in \eqref{eq:app:thm_3:list:G_G} finally yields \eqref{eq:app:thm_3:G_G_final} as shown on the next page.

Collecting terms from \eqref{eq:app:thm_3:A_n_A_l}, \eqref{eq:app:thm_3:A_n_G_l_k_final}, \eqref{eq:app:thm_3:G_G_final}, and substituting in \eqref{eq:app:thm_3:cross-terms}, Theorem~\ref{thm:e_r2} follows immediately.
This completes the proof.

\begin{figure*}
    \begin{align}
        &\mathbb{F}\left[G_{n,m}(t) G_{l,k}(t)\right] = \Ditilde^2 \alpha_n^2 \alpha_l^2 \left\lbrace \frac{1}{\left[\Ditilde (\alpha_n^2 + \alpha_l^2) + j \xi \right]\left(\Ditilde \alpha_n^2 + \Do \beta_k^2  + j \xi\right)\left[\Do(\beta_m^2 + \beta_k^2) + j \xi \right]} \right. \nonumber\\
        &\qquad \left. {}+ \frac{1}{\left[\Ditilde (\alpha_n^2 + \alpha_l^2) + j \xi \right]\left(\Do \beta_m^2 + \Ditilde \alpha_l^2 + j \xi \right)\left[\Do(\beta_m^2 + \beta_k^2) + j \xi \right]} \right\rbrace\label{eq:app:thm_3:G_G}
    \end{align}
    \hrule
\end{figure*}

\begin{figure*}
    \begin{align}
        &\Exp{G_{n,m}(t) G_{l,k}(t)} = \Ditilde^2 (n^2 + l^2) \pi^4 \zeta^{\gamma} \alpha (\alpha + 1) \nonumber\\ 
        & \qquad \cdot \bigintsss_{0}^{t} \bigintsss_{0}^{\theta} \left[ \frac{\exp\left(-\Do(\beta_m^2 (\theta - \tau) + \beta_k^2 \theta)\right)}{\left[\zeta + \Ditilde \pi^2 (n^2 (t + \tau - \theta) + l^2 (t - \theta)) \right]^{\gamma+2}} + \frac{\exp\left(-\Do(\beta_k^2 (\theta - \tau) + \beta_m^2 \theta)\right)}{\left[\zeta + \Ditilde \pi^2 (l^2 (t + \tau - \theta) + n^2 (t - \theta)) \right]^{\gamma+2}} \right] \dtau \dtheta\label{eq:app:thm_3:G_G_final}
    \end{align}
    \hrule
\end{figure*}

\fi

\renewcommand{\baselinestretch}{0.94}
\vspace*{-1.5mm}
\bibliographystyle{IEEEtran}
\bibliography{./IEEEabrv,references}

\end{document}